\documentclass[preprint,12pt]{elsarticle}

\usepackage{amssymb}

\biboptions{sort&compress}

\newcounter{bla}

\journal{Computer Physics Communications}

\begin{document}

\begin{frontmatter}



\title{GPUQT: An efficient linear-scaling quantum transport code fully implemented on graphics processing units}


\author[a]{Zheyong Fan\corref{author}}
\author[a]{Ville Vierimaa}
\author[a]{Ari Harju}

\cortext[author] {Corresponding author.\\\textit{E-mail address:} brucenju@gmail.com}
\address[a]{COMP Centre of Excellence and Helsinki Institute of Physics,
Department of Applied Physics, Aalto University, Helsinki, Finland}

\begin{abstract}
We present GPUQT, a quantum transport code fully implemented on graphics processing units. Using this code, one can obtain intrinsic electronic transport properties of large systems described by a real-space tight-binding Hamiltonian together with one or more types of disorder. The DC Kubo conductivity is represented as a time integral of the velocity auto-correlation or a time derivative of the mean square displacement. Linear scaling (with respect to the total number of orbitals in the system) computation time and memory usage are achieved by using various numerical techniques, including sparse matrix-vector multiplication, random phase approximation of trace, Chebyshev expansion of quantum evolution operator, and kernel polynomial method for quantum resolution operator. We describe the inputs and outputs of GPUQT and give two examples to demonstrate its usage, paying attention to the interpretations of the results.
\end{abstract}


\end{frontmatter}



{\bf PROGRAM SUMMARY}

\begin{small}
\noindent
{\em Manuscript Title:} GPUQT: An efficient linear-scaling quantum transport code fully implemented on graphics processing units                                       \\
{\em Authors:}
Zheyong Fan, Ville Vierimaa, and Ari Harju                    \\
{\em Program Title:}    GPUQT                                  \\
{\em Journal Reference:}                                      \\
{\em Catalogue identifier:}                                   \\
{\em Licensing provisions:}                                   \\
{\em Programming language: }   CUDA                            \\
{\em Computer:} Architectures with CUDA-enabled NVIDIA GPUs with compute capability 2.0 or higher. \\
{\em Operating system:}    Linux.                   \\
{\em RAM:} Needs about 1 to 10 GB device memory and less CPU memory, depending on the size of simulated system.  \\
{\em Number of processors used:}   One CPU processor and one GPU card.                 
{\em Classification:}   7.9 	Transport properties.           \\
{\em Nature of problem:} \\
Obtain intrinsic electronic transport properties of large systems described by real-space tight-binding Hamiltonians.\\
{\em Solution method:} \\
The DC conductivity is represented as a time integral of the velocity auto-correlation (VAC) or a time derivative of the mean square displacement (MSD). The calculations achieve linear scaling (with respect to the number of orbitals in the system) computation time and memory usage by using various numerical techniques, including sparse matrix-vector multiplication, random phase approximation of trace, Chebyshev expansion of quantum evolution operator, and kernel polynomial method for quantum resolution operator.\\
{\em Restrictions:} \\
The number of orbitals is restricted to about 20 million due to the limited amount of device memory in current GPUs.\\
{\em Running time:} About 3 minutes (using a Tesla K40 GPU card) for both examples provided.\\

\end{small}


\section{Introduction}

Electrical current can be either viewed as \cite{diventra2008} a consequence of an applied electric field as in the classical Boltzmann formalism and the quantum Kubo formalism, or as transmission of charge carriers as in the Landauer-B\"{u}ttiker formalism. The Landauer-B\"{u}ttiker formalism, or more generally the non-equilibrium Green's function formalism, is very versatile and has become the standard method for quantum transport simulations from nanoscale to mesoscale \cite{datta1995,ferry1997,huang2008,diventra2008,nazarov2009}. Combined with the recursive Green's function technique \cite{lewenkopf2013}, the Landauer-B\"{u}ttiker formalism can be used to efficiently simulate relatively narrow systems. An efficient and flexible open source code, Kwant \cite{groth2014}, is available for quantum transport simulations based on tight-binding models. As matrix inversion is at the heart of the recursive Green's function technique, the computational cost generally scales cubically with respect to the width of the system, which severely restricts the application of the method to realistically large 2D and 3D systems. To study large systems, linear-scaling computational cost is desirable. Fortunately, such a linear-scaling method has been developed in the Kubo formalism \cite{mayou1988,mayou1995,roche1997,roche1999,triozon2002,triozon2004} and has found a lot of applications, especially in quasi-1D and 2D materials \cite{markussen2006,ishii2010,ortmann2011,leconte2011,lherbier2012,cresti2013,
laissardiere2013,fan2014cpc,uppstu2014,fan2014prb,fan2015,ervasti2015,
fan2017}. For a review, see Ref. \cite{book_roche}.

Recently, we have made an efficient GPU implementation  \cite{fan2014cpc} of this method. Here, we present our GPU code, which we call GPUQT, and use a few examples to illustrate its usage.
GPUQT is fully implemented on the GPU using the CUDA toolkit \cite{cuda}. Using a single modern graphics card such as Tesla K40, the speedup factor achieved by GPUQT over a serial CPU implementation ranges from one to two orders of magnitude, depending on the problem. Usually, a higher speedup factor can be obtained in a problem with a denser Hamiltonian due to the higher arithmetic intensity, a measure of the amount of floating-point operations relative to the amount of memory accesses required to support those operations. Using GPUQT, one can easily simulate tight-binding systems with millions of sites. To use GPUQT, one has to define a simulation model by specifying the Hamiltonian and current (velocity) operators. The Hamiltonian should be defined in real space and be relatively sparse. As this method is not very suitable for studying ballistic transport properties \cite{fan2014cpc,markussen2006}, studied system should also contain one or more types of disorder such that a diffusive regime can be reached within a reasonable computation time.

This paper is organized as follows.
We first present the basic theoretical formalisms underlying GPUQT in section \ref{section:theory} and then
discuss the numerical techniques crucial for achieving linear scaling in section \ref{section:numerical}. In section \ref{section:usage}, we describe the overall structures of the GPUQT package and 
specifications in the inputs and outputs. Two examples are presented in section \ref{section:examples} to illustrate the usage of GPUQT. Section \ref{section:summary} summarizes and concludes.

\section{Theoretical formalisms\label{section:theory}}

\subsection{The Kubo-Greenwood formula for DC conductivity}

The Kubo-Greenwood formula \cite{kubo1957,greenwood1958} for DC diagonal conductivity
$\sigma^{\rm KG}(E)$ as a function of the Fermi energy $E$
at zero temperature is
\begin{equation}
\sigma^{\rm KG}(E) = \frac{2 \pi \hbar e^2}{\Omega}
            \textmd{Tr}\left[V \delta(E-H) V \delta(E-H)\right],
\end{equation}
where $\hbar$ is the reduced Planck constant, $e$ is the electron charge,
$\Omega$ is the system volume, $V$ is the velocity operator
in the  transport direction, $H$ is the Hamiltonian of the system, and
Tr denotes the trace. Spin degeneracy is included by the factor 2 in the formula. Linear-scaling evaluation of the Kubo-Greenwood conductivity has been studied early by Thouless and Kirkpatrick \cite{thouless1981}, later by Mayou\cite{mayou1988}, and recently by Ferreira and Mucciolo \cite{ferreira2015}.

\subsection{Velocity autocorrelation and mean square displacement}

In GPUQT, we do not directly calculate the Kubo-Greenwood conductivity $\sigma^{\rm KG}(E)$. Instead, we first calculate one of the two correlation functions, the velocity autocorrelation (VAC) or the mean square displacement (MSD). Both correlation functions are a function of the correlation time $t$.

By Fourier transforming one of the $\delta$ functions in the above formula, one can express the running electrical conductivity $\sigma^{\rm VAC}(E, t)$
as a time integral of the VAC $C_{vv}(E, t)$,
\begin{equation}
\label{equation:REC_VAC}
\sigma^{\rm VAC}(E, t) = e^2 \rho(E) \int_0^{t} C_{vv}(E,t) d t;
\end{equation}
\begin{equation}
C_{vv}(E, t) = \frac{\frac{2}{\Omega} \textmd{Re}
 \left[ \textmd{Tr} \left[U(t) V\delta(E - H) U(t)^{\dagger}V\right] \right]}
              {\frac{2 }{\Omega}\textmd{Tr}\left[\delta(E-H)\right]};
\end{equation}
\begin{equation}
\rho(E) = \frac{2 }{\Omega}\textmd{Tr}\left[\delta(E-H)\right],
\end{equation}
where $V(t) = U^{\dagger}(t) V U(t) = e^{iHt/\hbar}Ve^{-iHt/\hbar}$
is the velocity operator in the Heisenberg representation,
and $\rho(E)$ the density of states (DOS). The calculation of the VAC using linear-scaling techniques was first done by Mayou and Khanna \cite{mayou1995}.

Equivalently, one can expresses the running electrical conductivity as a time-derivative of the mean square displacement (MSD) $\Delta X^2(E, t)$,
\begin{equation}
\label{equation:REC_MSD}
\sigma^{\rm MSD}(E, t) = e^2 \rho(E) \frac{d}{2 d t}
            \Delta X^2(E, t),
\end{equation}
\begin{equation}
\Delta X^2(E, t) =
\frac{ \frac{2}{\Omega} \textmd{Tr}\left[[X, U(t)]^{\dagger}\delta(E - H) [X, U(t)]\right]}
{\frac{2}{\Omega}\textmd{Tr}\left[\delta(E-H)\right]},
\end{equation}
where $X(t) = U^{\dagger}(t) X U(t)$ is the position operator in the
Heisenberg representation. This formalism was first proposed by Roche and Mayou \cite{roche1997,roche1999} and later refined by Triozon \textit{et al}. \cite{triozon2002,triozon2004}.

\subsection{Transport regimes}

The VAC and MSD contain information about the transport regimes. For example, the MSD changes from a quadratic to a linear function of the correlation time during a ballistic-to-diffusive transition and finally saturates, causing the absence of diffusion, if strong (Anderson) localization takes place.

If the transport is diffusive, the VAC usually decays exponentially,
\begin{equation}
C_{vv}(E, t) = v_x^2(E) e^{-t/\tau(E)},
\end{equation}
where $\tau(E)$ is the relaxation time. Then we get the semiclassical conductivity:
\begin{equation}
\label{equation:sigma_sc}
\sigma_{sc}(E) = e^2 \rho(E) v_x^2(E) \tau(E).
\end{equation}
The product of the velocity and the scattering time is the mean free path \begin{equation}
\label{equation:mfp}
\lambda(E)=v_x(E) \tau(E). 
\end{equation}
Multiplying this with the velocity gives the diffusivity, $D(E)=v_x^2(E) \tau(E)$, which, in terms of the MSD, can be understood as an Einstein relation \cite{roche1997,roche1999}:
\begin{equation}
D(E) =\frac{1}{2}  
\lim_{t \rightarrow \infty} \frac{d}{dt} \Delta X^2(E, t) 
\approx \lim_{t \rightarrow \infty} \frac{\Delta X^2(E, t)}{2t}.
\end{equation}

When localization/anti-localization takes place, one usually needs to carefully analyse the behaviour of the running conductivity $\sigma(E,t)$.
Quantitative analysis of localization can be facilitated by the use of the MSD. The square root of the MSD,
\begin{equation}
\label{equation:L}
L(E, t) = 2 \sqrt{\Delta X^2(E,t)},
\end{equation}
serves as a good estimation of the length \cite{leconte2011,lherbier2012,fan2014cpc,uppstu2014,fan2014prb},
up to which the electrons propagate. Using this, we can express the running conductivity as a function of the propagating length, $\sigma(E,L)$, rather than the correlation time. In the strongly localized regime, the propogating length will finally saturate and the saturated value is found to be directly related to the localization length \cite{uppstu2014,fan2014prb},
\begin{equation}
\label{equation:xi}
\xi(E) =\lim_{t \rightarrow \infty} \frac{L(E,t)}{2\pi}.
\end{equation}

\section{Linear-scaling techniques\label{section:numerical}}

Based on the above theoretical formalisms, we see that
the quantities that need to be calculated are $\rho (E)$,
$\rho (E) C_{vv}(E, t)$, and $\rho (E) \Delta X^2(E, t)$. The running conductivity $\sigma (E,t)$ can then be
obtained either by a time integral of $\rho (E) C_{vv}(E, t)$, or a time derivative of $\rho (E) \Delta X^2(E, t)$. There are a few linear-scaling techniques which work together to achieve linear-scaling computation time and memory usage in the calculations of these quantities.

\subsection{Linear-scaling evaluation of the trace}

The first approximation is to use a random vector $|\phi\rangle$
to evaluate the trace \cite{weisse2006}:
\begin{equation}
 \textmd{Tr}\left[ A \right] \approx \langle \phi | A |\phi \rangle,
\end{equation}
where $A$ is an arbitrary $N \times N$ matrix operator, and $|\phi\rangle$
is normalized to $N$, $\langle \phi | \phi \rangle = N$.
The error introduced by this approximation decreases with increasing $N$, scaling as $\sim 1/\sqrt{N}$ \cite{weisse2006}.
For a given $N$, the accuracy can also be increased by taking average over
independent random vectors. The introduction of the random vector is crucial to achieving linear-scaling, because the major computation will be sparse matrix-vector multiplication, which scales linearly with respect to the system size.

With this approximation, we have
\begin{equation}
\label{equation:trace_DOS}
 \rho(E) \approx \frac{2}{\Omega} \langle \phi | \delta(E - H) |\phi \rangle;
\end{equation}
\begin{equation}
\label{equation:trace_VAC}
 \rho (E) C_{vv}(E, t) \approx
 \frac{2}{\Omega} \textmd{Re}
 \left[ \langle\phi|U(t)V \delta(E - H) U(t)^{\dagger}V|\phi \rangle \right];
\end{equation}
\begin{equation}
\label{equation:trace_MSD}
 \rho (E) \Delta X^2(E, t) \approx
 \frac{2}{\Omega}
 \langle\phi|[X, U(t)]^{\dagger} \delta(E - H)[X, U(t)]|\phi \rangle.
\end{equation}

\subsection{Linear-scaling evaluation of the time evolution}

Both the VAC and MSD formalisms involve a time evolution operator $U(t)$, which is absent from the Kubo-Greenwood formula. After using the random vector approximation, we only need to evaluate the application of the time evolution operator on a vector rather than matrix exponential. The basic idea is to divide the total correlation time into a number of steps. For one time step $\Delta t$ (the time steps need not to be uniform), we have the following Chebyshev polynomial expansions \cite{ezer1984,fehske2009,fan2014cpc}:
\begin{equation}
\label{equation:cheb_1}
 U(\pm \Delta t) |\psi\rangle
 \approx \sum_{m=0}^{N_p-1} (2-\delta_{0m}) (\mp i)^m
         J_m\left( \frac{\widetilde{\Delta t}}{\hbar} \right)
         T_m( \widetilde{H}) |\psi\rangle;
\end{equation}
\begin{equation}
\label{equation:cheb_2}
 [X, U(\Delta t)] |\psi\rangle
 \approx \sum_{m=0}^{N_p-1} (2-\delta_{0m}) (-i)^m
J_m\left( \frac{\widetilde{\Delta t}}{\hbar} \right)
[X, T_m(\widetilde{H})] |\psi\rangle,
\end{equation}
where $J_m$ is the $m$th order Bessel function of the first kind and $T_m$ is the $m$th order Chebyshev polynomial of the first kind.
Note that $T_m$ is defined in the interval $[-1,1]$ and the Hamiltonian and time step have to be scaled in the opposite way:
\begin{equation}
\widetilde{H}=H/\Delta E;
\end{equation}
\begin{equation}
\widetilde{\Delta t}= \Delta E \Delta t,
\end{equation}
where $\Delta E$ is sufficiently large such that the spectrum of the scaled Hamiltonian $\widetilde{H}$ lies within the interval $[-1,1]$. The Chebyshev polynomial expansions of the time evolution operators can be evaluated up to machine precision and the order of expansion $N_p$ needed for achieving this is proportional to the time interval. 

The above summations can be efficiently evaluated
by using the following  recurrence relations ($m \geq 2$)  ($T_m(\widetilde{H})$ is written as $T_m$ for simplicity) \cite{fan2014cpc}:
\begin{equation}
 T_m = 2 \widetilde{H} T_{m-1} - T_{m-2};
\end{equation}
\begin{equation}
[X, T_m] = 2[X, \widetilde{H}] T_{m-1} + 2\widetilde{H} [X, T_{m-1}] - [X, T_{m-2}];
\end{equation}
\begin{equation}
 T_0 = 1 \quad
 T_1 = \widetilde{H};
\end{equation}
\begin{equation}
 [X, T_0] = 0, \quad
 [X, T_1] = [X,\widetilde{H}].
\end{equation}

\subsection{Linear-scaling evaluation of the quantum resolution operator}

There are quite a few linear-scaling techniques for approximating the quantum resolution operator $\delta(E-H)$, including the Lanczos recursion method \cite{haydock1972,haydock1975}, the Fourier transform method  \cite{feit1982,hams2000}, and the kernel polynomial method \cite{weisse2006}. The Lanczos method is usually not as stable as the kernel polynomial method \cite{weisse2006}. In Ref. \cite{fan2014cpc}, it has been demonstrated that the Fourier transform method is not as efficient as the kernel polynomial method. We have thus implemented only the kernel polynomial method in GPUQT.

In the kernel polynomial method \cite{weisse2006}, the quantum resolution operator is approximated by a truncated Chebyshev polynomial expansion, and we can rewrite Eqs. (\ref{equation:trace_DOS}-\ref{equation:trace_MSD}) as
\begin{equation}
\label{equation:KPM_DOS}
 \rho(E) \approx \frac{2}{\pi \Omega \Delta E \sqrt{1-\widetilde{E}^2}}
 \sum_{n=0}^{N_m-1} g_n (2-\delta_{n0}) T_n(\widetilde{E})
     C_n^{\rm DOS};
\end{equation}
\begin{equation}
\label{equation:KPM_VAC}
 \rho (E) C_{vv}(E, t) \approx
 \frac{2}{\pi \Omega \Delta E \sqrt{1-\widetilde{E}^2}}
 \sum_{n=0}^{N_m-1} g_n (2-\delta_{n0})
     T_n(\widetilde{E}) C_n^{\rm VAC} (t);
\end{equation}
\begin{equation}
\label{equation:KPM_MSD}
 \rho (E) \Delta X^2(E, t) \approx
 \frac{2}{\pi \Omega \Delta E \sqrt{1-\widetilde{E}^2}}
 \sum_{n=0}^{N_m-1} g_n (2-\delta_{n0}) T_n(\widetilde{E})
     C_n^{\rm MSD} (t).
\end{equation}
Here, $C_n^{\rm DOS}$, $C_n^{\rm VAC}(t)$,
and $C_n^{\rm MSD}(t)$ are the Chebyshev moments:
\begin{equation}
\label{equation:C_n_DOS}
 C_n^{\rm DOS} \approx \langle \phi | T_n(\widetilde{H}) |\phi \rangle;
\end{equation}
\begin{equation}
\label{equation:C_n_VAC}
 C_n^{\rm VAC}(t) \approx \textmd{Re}
 \left[
 \langle\phi|U(t)V T_n(\widetilde{H}) U(t)^{\dagger}V|\phi \rangle
 \right];
\end{equation}
\begin{equation}
\label{equation:C_n_MSD}
 C_n^{\rm MSD}(t) \approx
 \langle\phi|[X, U(t)]^{\dagger} T_n(\widetilde{H}) [X, U(t)]|\phi \rangle.
\end{equation}
A kernel (damping factor) is applied before performing the
Chebyshev summation in order to suppress the Gibbs oscillations. For most applications, the Jackson damping \cite{weisse2006}
\begin{equation}
 g_n = \left(1 - n \alpha \right) \cos\left(\pi n \alpha \right)
     + \alpha \sin\left(\pi n \alpha \right) \cot\left(\pi \alpha\right),
\end{equation}
where $\alpha = 1/(N_m+1)$, is a good choice. The energy resolution achieved scales as $\delta \sim 1/N_m$ \cite{weisse2006}. Therefore, to achieve a finer energy resolution, one needs to use a larger $N_m$.

\section{Using GPUQT \label{section:usage}}

\subsection{Compile the code and run the examples}

After downloading and unpacking GPUQT, one can see two folders:  \verb"src" and \verb"examples". The folder \verb"src" contains all the source files and a \verb"makefile". The folder \verb"examples"
contains two sub-folders with names \verb"diffusive" and \verb"localized", both containing the files \verb"make_inputs.cpp" and \verb"plot_results.m".

To compile the code, simply go to the \verb"src" folder and type \verb"make" in the command line.
Upon finished, an executable called \verb"gpuqt" will be created in this folder.

Before running the examples, one has to first go to the \verb"diffusive" and \verb"localized" folders and compile (using e.g. \verb"g++") and run the \verb"make_inputs.cpp" code. This will create input files that are needed for running GPUQT.

Then, one needs an extra input file, which we call a ``driver input file'', to specify the path(s) of the folder(s) containing the input files. To run the two examples consecutively in a single job, this file should read
\begin{verbatim}
    examples/diffusive
    examples/localized
\end{verbatim}
Suppose that the ``driver input file'' is named as \verb"input.txt" and is in the \verb"examples" folder, one can run the examples using the following command:
\begin{verbatim}
    src/gpuqt examples/input.txt
\end{verbatim}

\subsection{Input files for GPUQT}

The input files are used to specify the Hamiltonian of a simulated system and some controlling parameters. All the input files for a simulation should be in a single folder.

In the tight-binding approximation, the Hamiltonian can be written as
\begin{equation}
 H = \sum_m\sum_n H_{mn} |m\rangle \langle n| 
 + \sum_m U_m |m\rangle \langle m|,
\end{equation}
where $H_{mn}$ is the hopping integral between sites (orbitals) $m$ and $n$ and $U_m$ is the on-site potential of site $m$. Similarly, the position and velocity operators can be expressed as
\begin{equation}
 X = \sum_{m} X_m |m\rangle \langle m|;
\end{equation}
\begin{equation}
 V = \frac{i}{\hbar} [H,X]
= \frac{i}{\hbar} \sum_m\sum_n
 (X_n-X_m) H_{mn} |m\rangle \langle n|.
\end{equation}
In the input files, one has to specify four sets of data: 1) the neighbor list structure that determines which hopping integrals are non-vanishing, 2) the non-vanishing hopping integrals, 3) the on-site potentials, and 4) the positions $X_m$ of the sites projected onto the transport direction.

\subsubsection{The neighbour.in input file}

This file specifies the topology of the problem using a neighbour list. This will be used to build the sparse Hamiltonian. The first line should have two integer numbers. The first number is the total number of sites in the simulated system. The second number is the maximum possible number of nonzero hopping integrals originated from a given site. For example, in a square lattice with nearest-neighbour hopping only, this number can be set as 4. Using a larger number than needed will waste memory. Starting from the second line, the $n$th line contains the number of neighbours and the indices of the neighbouring sites of the $(n-1)$th site.

\subsubsection{The hopping.in input file}

This is an optional input file, which contains the hopping integrals (the off-diagonal terms in the Hamiltonian). If this file is not prepared, GPUQT assumes that all the hopping integrals between pairs of neighbouring sites (specified in the neighbour.in file) are $-1$. 
The first line should be either the word \verb"real" or \verb"complex".
If the word is \verb"real", it means that all the hopping integrals are real numbers. Then, starting from the second line, the $n$th line contains the real hopping integrals between the $(n-1)$th site and its neighbouring sites, and the order should be consistent with that in the neighbour.in file. The file will look like this:
\begin{verbatim}
    real
    real_1 real_2 real_3 ...
    ...
\end{verbatim}
If the word is \verb"complex", it means that not all the hopping integrals are real numbers. Then, each real hopping integral as described above should be substituted by two real numbers, the real and imaginary parts of the complex hopping integral. The file will look like this:
\begin{verbatim}
    complex
    real_1 imag_1 real_2 imag_2 real_3 imag_3 ...
    ...
\end{verbatim}
The unit of energy is determined by the user. One should consistently use the same unit in other input files such as \verb"potential.in", \verb"energy.in" and \verb"para.in".

\subsubsection{The potential.in input file}

This is an optional input file, which contains the on-site potentials (the diagonal terms in the Hamiltonian). If this file is not prepared, GPUQT assumes that all the on-site potentials are zero. The $n$th line is the on-site potential of the $n$th site.

\subsubsection{The position.in input file}

This file specifies the coordinates of the sites in the simulated system. This will be used to build the velocity (current) operator. The first line should have two numbers, which are the length of the simulated system in the transport direction and the volume of the system. Be careful with periodic boundary conditions. For example, consider a $1~000 \times 1~000$ regular square lattice with a lattice constant of $a=1$, the length in the $x$ direction should be 1 000, even though the distance between a leftmost site and a rightmost site is only 999. Starting from the second line, the $n$th line is the position component of the $(n-1)$th site in the transport direction. The unit of length is determined by the user. One can either set the lattice constant to 1 or some values in unit of nm or \AA. What is important is to be consistent when reporting the results.

\subsubsection{The energy.in input file}

This file contains the energy points to be considered in the simulations. There is a single column in this file. The first line should be an integer, which is the number of energy points to be read in. Starting from the second line, the $n$th line contains the $(n-1)$th energy value. Note that the method is parallel in energy and using one thousand energy points takes roughly as much time as using a single energy point. 
Here is an example:
\begin{verbatim}
    601
    -3.00
    -2.99
    ...
    0.00
    ...
    3.00
\end{verbatim}
This file tells that there would be 601 energy points to be calculated, from $-3$ to $3$, with a spacing of $0.01$.

\subsubsection{The time$\textunderscore$step.in input file}

This file contains the time steps in the VAC and/or MSD calculations. There is a single column in this fie.
The first line should be an integer, which is the number of time steps to be read in. Starting from the second line, the $n$th line is the $(n-1)$th time step.
Here is an example:
\begin{verbatim}
    20
    1
    2
    ...
    19
    20
\end{verbatim}
This file tells that there would be 20 (non-uniform) time steps, from $t_0$ to 20 $t_0$, with a spacing of $t_0$. One should note that the data here are the time steps, not the cumulative times. The cumulative times for this example should be $t_0$, $3t_0$, $6t_0$, $10t_0$, $\cdots$. The unit of time, $t_0$ is fixed by the energy unit, as we set the reduced Planck constant to 1 in GPUQT. Suppose the unit of energy is $\gamma$, the unit of time is then $t_0=\hbar/\gamma$.

\subsubsection{The para.in input file}

This file contains some additional parameters to define the simulation.
In this input file, blank lines are ignored. Each non-empty line starts with a keyword possibly followed by one or more parameters. The valid keywords and their parameters are
\begin{enumerate}
\item \verb"calculate_vac"\\
This keyword does not need any parameter. If this keyword appears, the VAC will be calculated. Otherwise, the VAC will not be calculated.
\item \verb"calculate_msd"\\
This keyword does not need any parameter. If this keyword appears, the MSD will be calculated. Otherwise, the MSD will not be calculated. If both this the above keywords are absent, there is no need to prepare the \verb"time_step.in" input file.
\item \verb"number_of_random_vectors" $N_r$\\
This keyword needs one parameter, which is the number of random vectors $N_r$ used in the simulation. If this keyword is absent, the default value $N_r=1$ will be used. If you want to use 10 random vectors for a given problem, you can either set this number to 10, or set it to 1 and then run the simulation 10 times. Increasing $N_r$ can improve the accuracy of the results.
\item \verb"number_of_moments" $N_m$\\
This keyword needs one parameter, which is the number of Chebyshev moments $N_m$ used in the kernel polynomial method. If this keyword is absent, the default value $N_m=1000$ will be used. A larger $N_m$ gives a finer energy resolution and one usually needs to test the effects of this parameter.
\item \verb"energy_max" $\Delta E$ \\
This keyword needs one parameter, which is a scaling parameter $\Delta E$ used to scale the Hamiltonian. The scaled Hamiltonian $H/\Delta E$ must have all of its eigenvalues lying within the interval $[-1, 1]$. If this keyword is absent, the default value $\Delta E =10$ will be used. Using a value larger than needed will only effectively reduce the energy resolution, but using a value smaller than needed will cause big problems as this will lead to calculating the square roots of negative numbers.
\end{enumerate}

\subsection{Output files of GPUQT}

We now describe the data format of the output files produced by running GPUQT. We note that for all the output files, results from a new simulation will append to, rather than overwrite existing data.

\subsubsection{The dos.out file}

The $n$th column of this file corresponds to the value of $\rho(E_n)$ at the $n$th energy point $E_n$ specified in the \verb"energy.in" file. Each row corresponds to the results obtained by using one random vector. The unit of DOS is $1/\gamma/a^2$ in 2D and $1/\gamma/a^3$ in 3D, where $\gamma$ is the unit of energy and $a$ is the unit of length.

\subsubsection{The vac.out file}

The $n$th column of this file corresponds to the value of $\rho(E_n)C_{vv}(E_n,t)$ at the $n$th energy point $E_n$ specified in the \verb"energy.in" file. If the number of time steps specified in the \verb"time_step.in" file is $N_t$, the first $N_t$ rows correspond to the results obtained by using one random vector. Integrating this quantity with respect to time gives the running electrical conductivity. In 2D, the unit of conductivity is $e^2/\hbar$. In 3D, the unit is $e^2/\hbar/a$, where $a$ is the unit of length. As expected, the unit in 3D can be converted to S/cm.

\subsubsection{The msd.out file}

The $n$th column of this file corresponds to the value of $\rho(E_n) \Delta X^2(E_n,t)$ at the $n$th energy point $E_n$ specified in the \verb"energy.in" file. If the number of time steps specified in the \verb"time_step.in" file is $N_t$, the first $N_t$ rows correspond to the results obtained by using one random vector. Taking derivative of this quantity with respect to time and then dividing by 2 gives the running electrical conductivity.

\section{Examples \label{section:examples}}

In this section, we present two examples to illustrate the usage of GPUQT.
Although this method has been mostly used to study graphene-based materials \cite{ortmann2011,leconte2011,lherbier2012,cresti2013,
laissardiere2013,fan2014cpc,uppstu2014,fan2014prb,fan2015,ervasti2015,
fan2017}, any system with an appropriate real-space tight-binding Hamiltonian can be treated. Here, for pedagogical purposes, we consider the Anderson model of square lattice. The tight-binding Hamiltonian reads
\begin{equation}
H = - \sum_{\langle m,n\rangle} \gamma |m\rangle \langle n| + \sum_m U_m |m\rangle \langle m|,
\end{equation}
where $\langle m, n\rangle$ means a pair of sites which are nearest neighbors of each other. The on-site potentials $U_m$ take values uniformly distributed in $[-W/2,W/2]$, where $W$ is the strength of the Anderson disorder.

\subsection{The diffusive regime in a 2D lattice}

\begin{figure*}[ht]
  \centering
  \includegraphics[width=0.7\columnwidth]{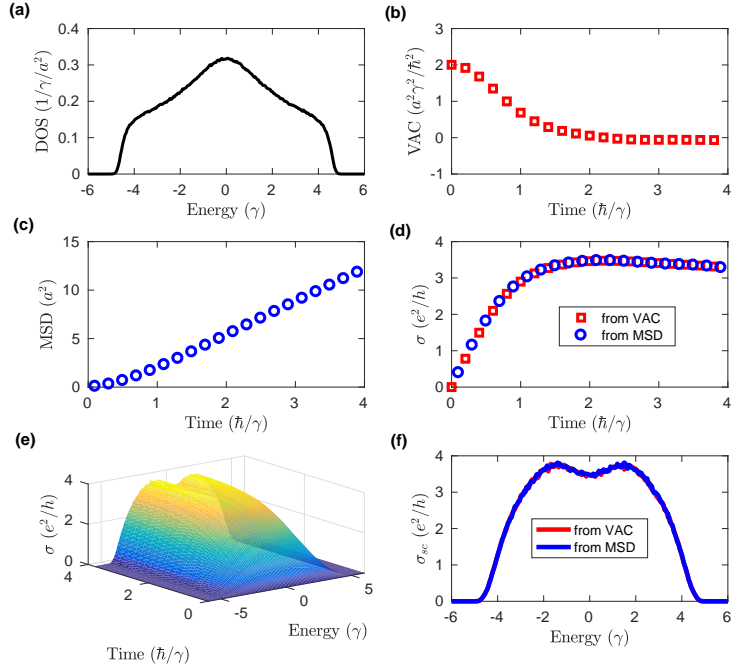}\\
  \caption{Results for a 2D square lattice with Anderson disorder of $W=4$. (a) Density of states (DOS) as a function of Fermi energy. (b) Velocity autocorrelation (VAC) and (c) mean square displacement (MSD) as a function of correlation time $t$ at energy $E=0$. (d) Running electrical conductivity at energy $E=0$ as a function of correlation time calculated from the VAC or MSD. (e) Surface plot of $\sigma^{\rm VAC}(E,t)$. (f) Semiclassical conductivity $\sigma_{sc}$ as a function of Fermi energy calculated from the VAC or MSD.}
  \label{fig:1}
\end{figure*}

In this example, we consider a $2000\times 2000$ square lattice with $W=4\gamma$, using periodic boundary conditions in both directions. This generally represents an effectively 2D system, although one usually needs to check possible finite-size effects. The input files can be generated by compiling and running the code \verb"make_inputs.cpp" prepared in the \verb"examples/diffusive" folder. Using $\gamma$ as the energy unit, all the nonzero hopping integrals have the value of $-1$. Therefore, the \verb"hopping.in" file is not needed. A uniform time step of $0.2\hbar/\gamma$ is used and the number of time steps is 20, which gives a total correlation time of $4\hbar/\gamma$. The energy points considered range from $-6\gamma$ to $6\gamma$, with an interval of $0.02\gamma$. 
The \verb"para.in" file reads:
\begin{verbatim}
    energy_max 6.1
    calculate_vac
    calculate_msd
\end{verbatim}
which means that $N_m=1000$ (default value), $\Delta E=6.1$, $N_r=1$ (default value), and both VAC and MSD will be calculated. 

After running this example, three output files, \verb"dos.out", \verb"vac.out", and \verb"msd.out" will be generated. Running the MATLAB script \verb"plot_results.m" prepared in the same folder will produce the graphs shown in Fig. \ref{fig:1}. The DOS in Fig. \ref{fig:1} (a) shows that the van Hove singularity at the band center in ordered 2D square lattice disappears in the presence of Anderson disorder. With increasing correlation time $t$, the VAC [Fig. \ref{fig:1} (b)] decays, while the MSD [Fig. \ref{fig:1} (c)] changes from a quadratic to a linear function, both indicating a ballistic-to-diffusive transition. The running conductivities calculated from the VAC and the MSD, $\sigma^{\rm VAC}(E,t)$ and $\sigma^{\rm MSD}(E,t)$, are equivalent to each other, as demonstrated in Fig. \ref{fig:1} (d) for the band center $E=0$. Up to a correlation time of $4\hbar/\gamma$, diffusive transport has been achieved for all the energy points, as can be seen from Fig. \ref{fig:1} (e). Further increasing the correlation time (propagating length) will bring the transport into the localized regime, leading to decreasing running conductivity. Therefore, it is reasonable to take the maximum value (there are exceptions \cite{fan2014prb}, though) at each energy $E$ as the semiclassical conductivity $\sigma_{sc}(E,t)$, which is shown in Fig. \ref{fig:1} (f). From the semiclassical conductivity and the electron group velocity, which is simply square root of the VAC at zero correlation time, one can obtain the relaxation time and the mean free path using Eqs. (\ref{equation:sigma_sc}) and (\ref{equation:mfp}).

\subsection{The localized regime in a 1D chain}

\begin{figure*}[ht]
  \centering
  \includegraphics[width=0.7\columnwidth]{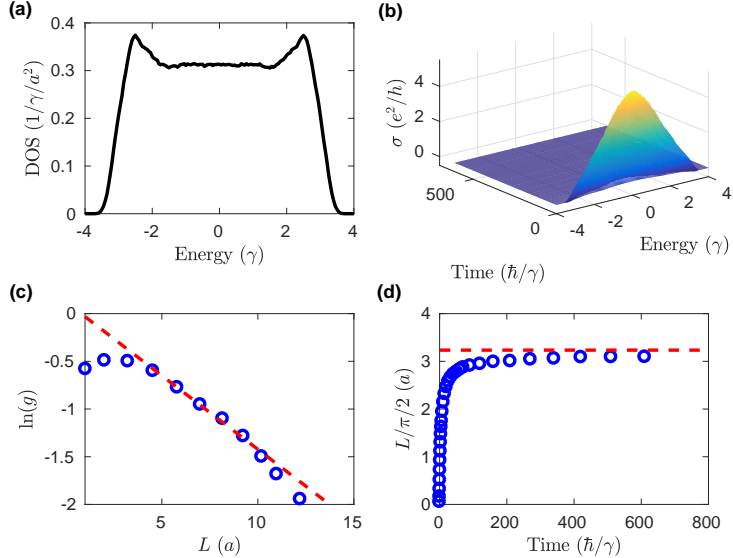}\\
  \caption{Results for a 1D chain with Anderson disorder of $W=4$. (a) Density of states (DOS) as a function of Fermi energy. (b) Surface plot of $\sigma^{\rm MSD}(E,t)$. (c) Logarithmic of the dimensionless conductance $\ln[g(E,L)]$ as a function of the propagating length $L$ at $E=0$. (d) $L(E,t)/2\pi$ as a function of correlation time $t$ at $E=0$. The dashed line in (c) is a fit using Eq. (\ref{equation:g_L}) and the dashed line in (d) indicates the localization length $\xi$ extracted from the fit.}
  \label{fig:2}
\end{figure*}

In this example, a 1D disordered chain of length $4~000~000$ and with $W=4\gamma$ is considered, using periodic boundary conditions. The input files can be generated by compiling and running the code \verb"make_inputs.cpp" prepared in the \verb"examples/localized" folder. Again, all the nonzero hopping integrals have the value of $-1$ in unit of $\gamma$ and the \verb"hopping.in" file is not needed. The energy points considered range from $-4\gamma$ to $4\gamma$, with an interval of $0.02\gamma$. As explained in Ref. \cite{fan2014cpc}, the VAC formalism is not as practical as the MSD formalism in the localized regime. Also, the time steps do not need to be uniform. The \verb"time_step.in" for this example reads:
\begin{verbatim}
    30
    0.1
    0.2
    ...
    1.0
    1
    2
    ...
    10
    10
    20
    ...
    100
\end{verbatim}
That is, we use larger and larger time steps, which can capture both the ballistic-to-diffusive transition and the diffusive-to-localized transition. The \verb"para.in" file reads:
\begin{verbatim}
    number_of_random_vectors 2
    number_of_moments        500
    energy_max               4.1
    calculate_msd
\end{verbatim}
which means that $N_m=500$, $\Delta E=4.1$, $N_r=2$, and only MSD will be calculated. 

After obtaining the output files by running GPUQT, one can run the MATLAB script \verb"plot_results.m" prepared in the same folder to get the graphs shown in Fig. \ref{fig:2}. The DOS in Fig. \ref{fig:2} (a) shows smoothed Hove singularities at the band edges in 1D chain. The running conductivity in Fig. \ref{fig:2} (b) clearly shows a maximum at each energy corresponding to $\sigma_{sc}(E,t)$ and the vanishing of conductivity at large correlation time due to Anderson localization. As discussed in Ref. \cite{uppstu2014}, the conductance $g(E,L)=\sigma(E,L) A/L$ ($A$ is the width of the system, which is 1 here) decays exponentially in an appropriate range of the propagating length $L(E)$:
\begin{equation}
\label{equation:g_L}
g(E,L) \sim e^{-L(E)/2\xi(E)},
\end{equation}
as shown in Fig. \ref{fig:2}(c). The localization length $\xi(E)$ is equivalent to that defined in Eq. (\ref{equation:xi}), as can be seen from Fig. \ref{fig:2}(d).

\section{Summary and Conclusions \label{section:summary}}

We have presented GPUQT, an efficient CUDA code suitable for studying intrinsic quantum transport properties of large systems described by real-space tight-binding Hamiltonians. Although we have used 2D square lattice and 1D chain with Anderson disorder to illustrate the usage, the inputs to GPUQT are made as general as possible such that many more realistic problems can be studied. The current version is only able to calculate the diagonal conductivity. Hall conductivity \cite{garcia2015} and spin relaxation time \cite{vantuan2016,vierimaa2017} can also be calculated within the same framework and their implementation will be included in a future version. The code and its updating can be accessed from GitHub \cite{my_github}.





\bibliographystyle{elsarticle-num}



\end{document}